\documentclass[structabstract]{aa}
\usepackage{amsmath,amsfonts,amssymb,amsxtra} 
\usepackage{txfonts}
\usepackage{graphicx}
\usepackage{natbib}
\usepackage{float}
\usepackage{tabularx}
\usepackage[figuresright]{rotating}

\usepackage{color}

\newcommand{\Msun}{${M_\odot}$}

\newcommand{\lsim}{\mathrel{\hbox{\rlap{\lower.55ex \hbox {$\sim$}}
 \kern-.3em \raise.4ex \hbox{$<$}}}}
\newcommand{\gsim}{\mathrel{\hbox{\rlap{\lower.55ex \hbox {$\sim$}}
 \kern-.3em \raise.4ex \hbox{$>$}}}}

\begin{document}

\title{On the nature and detectability of Type Ib/c supernova progenitors} 
\author{S.-C. Yoon\inst{1}  \and G. Gr\"afener\inst{2}  \and  J. S. Vink\inst{2} \and A. Kozyreva\inst{1} \and R. G. Izzard\inst{1}}

\institute {
Argelander-Institut f\"ur Astronomie der Universit\"at Bonn, Auf dem H\"ugel 71, 53121 Bonn, Germany\\
\email{scyoon@astro.uni-bonn.de}
\and
Armagh Observatory, College Hill, Armagh, BT61 9DG, United Kingdom
}

\date{Received:  / Accepted: }

\abstract
{The progenitors of many Type II supernovae have been observationally identified but 
the search for Type Ibc supernova (SN Ibc) progenitors has thus far been unsuccessful, despite the 
expectation that they are luminous Wolf-Rayet (WR) stars. 
}
{We investigate how the evolution of massive helium stars affects their visual 
appearances, and discuss the implications for the detectability of SN Ibc progenitors. 
}
{Evolutionary models of massive helium stars are analysed and their properties compared to Galactic WR stars. 
}
{Massive WR stars that rapidly lose their helium envelopes through stellar-wind
mass-loss end their lives when their effective temperatures -- related to their
hydrostatic surfaces -- exceed about 150kK.
%
At their pre-supernova stage, their surface properties 
resemble those of hot Galactic WR stars of WO
sub-type. These are visually faint with narrow-band visual magnitudes $M_v = -1.5 \cdots -2.5$, despite 
their high bolometric luminosities ($\log L/L_\odot = 5.6 \cdots 5.7$),   
compared to the bulk of Galactic WR stars ($M_v < -4$). 
In contrast, relatively low-mass helium stars that
retain a thick helium envelope appear fairly bright in optical bands,
depending on the final masses and the history of the envelope expansion during
the late evolutionary stages.}
{We conclude that SNe Ibc observations 
have so far not provided strong constraints on 
progenitor bolometric luminosities and masses, even with the deepest searches. 
We also argue that Ic progenitors are more challenging to identify 
than Ib progenitors in any optical images.
}


\maketitle


\section{Introduction}\label{sect:intro}

Direct detections of the progenitor stars of supernovae discovered in the nearby
Universe provide one of the most stringent constraints on  stellar
evolution theory. Since the unambiguous identification of the progenitor of the
supernova 1987A, observers have identified progenitor stars of numerous Type IIP
supernovae and several Type IIb and IIn supernovae in pre-supernova images
\citep[see][for a review]{Smartt09}.  However, 
despite their relevance to Galactic chemical evolution, the progenitor stars 
of supernovae (SNe) Ibc remain as yet elusive. 

For decades, there had been a widely held belief that SNe Ibc progenitors are massive
Wolf-Rayet (WR) stars,  formed either through stellar-wind mass-loss or Roche-lobe overflow
in a binary system. Also the identification of broad SN Ic features in the vicinity
of several long GRBs \citep{Galama98, Hjorth03, Stanek03} for which the
progenitors were also suggested to be massive WR stars \citep{Woosley93} added
further evidence to this assertion. An alternative to the massive WR scenario
is that of lower mass binaries \citep[e.g.][]{Podsiadlowski92, Wellstein99, Eldridge08, Yoon10}. 

Although many previous searches for SNe Ibc progenitors
were hampered by insufficient detection limits, WR stars with $\log
L/L_\odot \gsim 5.3$ could have been identified as the progenitors of the
Type Ib supernova 2000ds~\citep{Maund05a}, and the Type Ic supernovae
2004gt~\citep{Maund05b} and 2002ap~\citep{Crockett07}. 
On the other hand, the most recent stellar evolution models predict that WR
stars originating from massive single stars with initial masses higher than
about 30~\Msun{} have bolometric luminosities of $\log L/L_\odot \gsim
5.4$ at the pre-supernova stage~\citep{Meynet03, Crockett07, Georgy12}.  New
detection limits, e.g., on SNe 2002ap, have mostly been interpreted as
evidence against the massive WR scenario, but in favour of progenitors of lower mass binaries instead
and largely dismissing the option of massive WR stars
\citep[e.g.,][]{Crockett07, Smartt09, Yoon10}.  

However, in these interpretations the crucial final stages of massive-star
evolution are not properly accounted for, as we argue in the
following.  In this paper, we present the seemingly counter-intuitive
idea that the more massive -- and bolometrically more luminous -- single
WR progenitors are more challenging to detect than low-mass binaries.

\section{Massive Wolf-Rayet stars: Models and observations}\label{sect:wr}

One of the most crucial questions regarding the observational identification of
massive WR stars as SN Ibc progenitors is this: which bolometric correction
(BC) should be applied to WR stars at the pre-supernova stage?  Because the
optical luminosities of WR stars are critically affected by optically thick WR
winds, the simple assumption of black-body radiation should not be applied even
for a crude approximation \citep[cf.][]{Crowther07}. 


\citet{Maund05a} and \citet{Maund05b} calculated BCs for WR stars using the
Potsdam grid of synthetic WR spectra~\citep{Hamann04,Graefener02} and obtained
the detection limits for different SNe Ibc progenitors in terms of bolometric
luminosities.  These detection limits were then compared to the evolutionary
tracks of massive stars given by \citet{Meynet94} to constrain the SN Ibc
progenitor masses. The caveat here is that \citet{Meynet94} applied a correction
to the radii of their WR star models to take into account the radial extension of the star
due to an optically thick wind, following \citet{Langer89} and \citet{Maeder90}.  This
correction is particularly large as they employed high WR mass-loss rates for
which the effect of wind clumping was not considered \citep[cf.][]{Hamann98}.  
As a consequence, the evolutionary tracks of their WR
star models have  surface temperatures of  $\log T \simeq 4.8$, and Maund et al. applied their BCs 
accordingly only up
to the quoted value.  However, the stellar temperatures $T_\star$
of the Potsdam models do not refer to the values at the photosphere within the
extended optically thick wind, but to those at large optical depths close to the
hydrostatic stellar surface.  This inconsistency introduced a
large  underestimate of the BCs applied to WR stars at the pre-supernova
stage, because stellar evolution models predict that massive WR stars with masses $M \gsim
10$§~\Msun{} end their lives with hydrostatic surface temperatures well
above $\log T_\star  = 5.0$ (see discussion below).  On the other hand,
\citet{Crockett07} simply employed a constant BC of -4.5 for the entire
temperature range of WR stars, following \citet{Smith89}. More recent analyses of WR
stars, however, seriously question the validity of this approach, as discussed
below. 

\begin{figure}
\centering
\includegraphics[width=1.0\columnwidth]{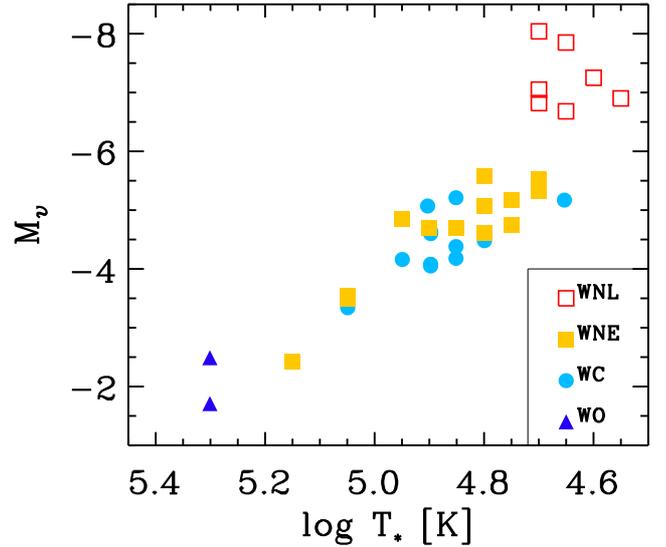}
\caption{Absolute visual magnitudes of Galactic Wolf-Rayet stars with known
distances of different spectral types (WNL, WNE, WC and WO)
as a function of surface temperatures. The data were taken from \citet{Hamann06}
and \citet{Sander12}. 
}
\label{fig:mv}
\end{figure}

\begin{figure}
\centering
\includegraphics[width=1.0\columnwidth]{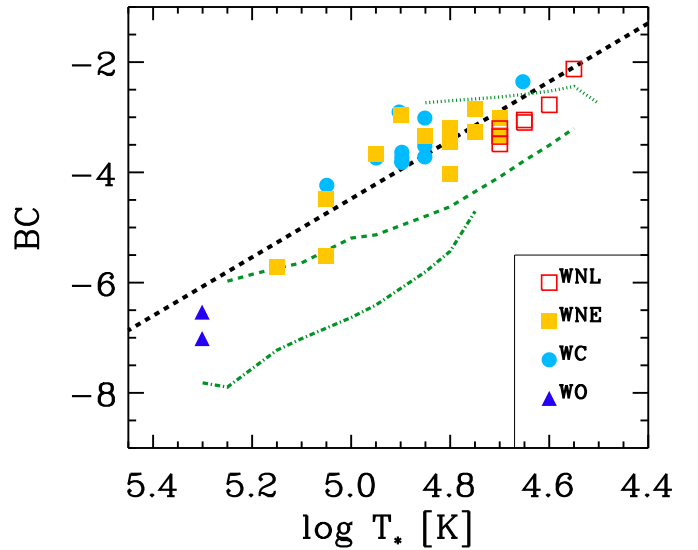}
\caption{Bolometric corrections (BCs) for Galactic WR stars, calculated
from the data given in Fig.~\ref{fig:mv},  as 
a function of surface temperatures. The thick dashed line gives the best fit to the data.  
The thin lines denote the BCs calculated by \citet{Maund05b}
from the Potsdam WR-star synthetic spectra of \citet{Graefener02}, 
for the minimum (dotted line), median (dashed line), and maximum (dot-dashed line)
transformed radii $R_t$ (see Maund et al. for details). 
}
\label{fig:bc}
\end{figure}

\citet{Hamann06} and \citet{Sander12} (hereafter, the Potsdam group) 
provide a homogeneous set of data on the properties of Galactic WR stars of
different spectral types (WNL, WNE, WC, and WO).  Fig.~\ref{fig:mv} shows the
absolute narrow-band visual magnitudes $M_v$\footnote{Throughout this paper, 
$M_v$ represents an absolute magnitude in the narrow visual-band ($3900\lsim \lambda \lsim 4500$).} 
of the Galactic WR stars with known
distances, as given by the Potsdam group. We note the strong
temperature dependence of $M_v$: WR stars tend to be visually dimmer at 
higher temperatures.  
The corresponding BCs are presented in Fig.~\ref{fig:bc}, 
for which the best linear fit is given by
\footnote{
In stellar evolution models, the hydrostatic surface temperature
corresponds to the effective temperature at the surface boundary for which
radial extension of the star due to optically thick WR winds is not considered. 
The empirical stellar
temperatures $T_\star$
in the Potsdam group data were estimated at large optical depth
($\tau_{\rm Ross}=20$) in
the WR wind models \citep{Sander12}  and thus nearly resemble the
hydrostatic values. However, as the hydrostatic layers are not
directly observable, the empirical $T_\star$ rely on the adopted wind
structure. For the stars with the strongest winds, we estimate a
possible systematic uncertainty of up to 0.1 in $\log T_\star$ caused by
this effect.
}
\begin{equation}
\mathrm{BC} = 22.053 - 5.306 \log T_\star~.
\end{equation}
To compute $BC = M_\mathrm{bol} - M_V$, we adopt $M_v - M_V = 0.75$ to correct for the
effect of strong emission lines on the visual broad-band magnitude $M_V$
\citep[cf.][]{Crockett07}. 
The BCs calculated by \citet{Maund05b} using the Potsdam WR grid
are plotted in the same figure for comparison.  Here again, the
figure reveals that BCs for WR stars are a sensitive function of temperature,
and that a more negative BC should be applied at a
higher temperature. The synthetic spectra from the Potsdam WR models
agree reasonably well with observations, confirming the
temperature dependence of BCs.

\begin{figure}
\centering
\includegraphics[width=1.0\columnwidth]{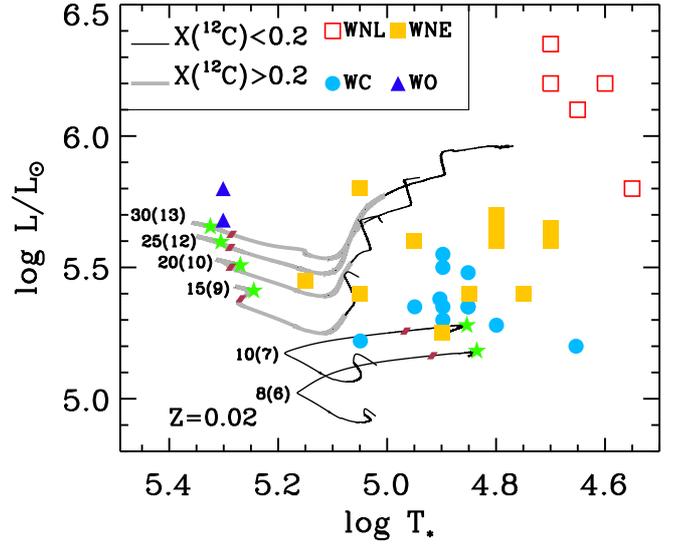}
\caption{Evolutionary tracks of pure helium-star models at solar metallicity
in the HR diagram, for six different initial masses (8, 10, 15, 20, 25, and 30~\Msun), as indicated by the labels. 
The numbers in the parentheses denote the corresponding final masses. 
The positions where the
surface mass-fraction of carbon is larger than 0.2 are marked by thick
grey lines. 
The star symbol marks the end point of the evolution, which is  the beginning of 
core oxygen-burning that occurs a few years before core collapse, for each evolutionary sequence, while
the tilted rectangle denotes the position at 1000 yrs before the supernova explosion. 
The positions of Galactic WR stars with known distances given by \citet{Hamann06} 
and \citet{Sander12} are marked by open squares (WNL), filled squares (WNE),
filled circles (WC), and triangles (WO). 
}
\label{fig:hr}
\end{figure}

To determine the implications of this finding for SN Ibc progenitors, we 
show a Hertzsprung-Russell (HR) diagram 
of evolutionary tracks of single massive helium stars of solar metallicity
in Fig.~\ref{fig:hr}. Our models were calculated with the
hydrodynamic stellar-evolution code described in \citet{Yoon10} using the WR
mass-loss rates from \citet{Nugis00}. The figure shows that helium stars with
initial masses higher than about 15~\Msun{} become gradually hotter. This is
because they progressively lose their helium-rich envelopes in a stellar-wind, exposing
the carbon-oxygen cores that gradually contract.  The corresponding final
masses are higher than about 9.0~\Msun{} and  the surface temperatures
increase to $\log T_\star \simeq 5.3$ at the end of their evolution.  We note that the recent single
star-models of the Geneva group predict similar surface temperatures for
WR stars of $\log L/L_\odot \gsim 5.3$ at the pre-supernova stage
\citep{Georgy12}. The question to be addressed now is which bolometric
correction should be applied to such WR stars with
$\log T_\star \gsim 5.3$. 

Fig.~\ref{fig:hr} shows the well-known discrepancy between the
predicted surface temperatures of WR stars and the observationally derived
values:  the  helium-star models generally give higher temperatures
than those of observed WR stars.  \citet{Graefener12} suggested
that the observationally implied large radii of many Galactic WR stars are 
explained by envelope inflation caused by the iron opacity peak at 
$\log T \simeq 5.18$ and the effect of density inhomogeneities
(clumping) in the inflated layers.  These inflated envelopes usually contain
a strong density inversion, as is often observed in stellar
models~\citep{Ishii99, Petrovic06}.   
This density inversion is also observed in 
our helium-star models presented
in Fig.~\ref{fig:hr},  but the degree of inflation in our models is much weaker than in
the \citet{Graefener12} models because we do not consider the effect of clumping.  
Interestingly, given the key role of the iron peak, this
inflation may no longer occur when the surface temperature of a star
approaches or becomes hotter than the iron peak temperature ($\log T \simeq 5.18$).
Our helium-star models indeed show that the density inversion in the outermost
layers, a sign of inflation caused by the iron peak, disappears when $\log T_\star
\gsim 5.07$.  This implies that, although massive WR stars  have
inflated envelopes for most of their lifetimes, the inflation disappears as
they lose most of their helium envelopes and their cores  become more 
compact during the late evolutionary stages.  \citet{Graefener05} showed
that these compact stars form hot WR-type winds driven by the radiative
force on the iron peak opacities.  This may explain why the envelopes of these
stars do not inflate and the positions of early-type WC and WO stars agree
well with the model predictions, while WR stars of later spectral subtypes have
much cooler temperatures.

Although we cannot directly apply the BCs of Fig.~\ref{fig:bc} to
the stellar evolution models, we can draw the following conclusion:
\emph{massive WR stars experiencing strong mass-loss end their lives 
at their hottest point ($\log T_\star  \simeq  5.3$).
Therefore, SN Ibc progenitors with final masses higher than about 9~\Msun{}
resemble the WO stars with $\log T_\star \simeq 5.2$ and $M_v = -1.5 \cdots 
-2.5$ (Fig.~\ref{fig:mv}), rather than the majority of observed WR stars
($\log T_\star \lsim 5$ and $M_v \lsim -4$).}  They are relatively faint
in optical bands compared to the majority of observed Galactic WR stars, and
we cannot exclude massive WR stars of $\log L/L_\odot > 5.4$ as the
progenitors of SNe 2000ds, 2004gt and 2002ap, in contrast to the previous
conclusions of \citet{Maund05a}, \citet{Maund05b}, and \citet{Crockett07}.

On the other hand, the helium envelopes in less-massive helium stars ($M  =$ 8
and 10~\Msun{}  Fig.~\ref{fig:hr}) are not greatly stripped off and  carbon is
not enriched at the surface.  These stars therefore rapidly expand during the late
evolutionary stages beyond core helium-burning because of the mirror effect
\citep[cf.][]{Kippenhahn90}.  The surface temperature decreases from $\log T_\star
\simeq 5.2$ at core helium-exhaustion to $\log T_\star \simeq 4.9$ at core oxygen-burning, accordingly.  
Although the bolometric luminosities of these SN Ibc
progenitor models ($\log L/L_\odot \approx 5.2$) are lower than those of more
massive ones ($\log L/L_\odot \gsim 5.4$), much lower temperatures 
at the pre-supernova stage imply that
they should appear more luminous in optical bands.  Given that WR stars
with $\log T_\star < 5.0$ and $\log L/L_\odot < 5.3$ in the Potsdam sample have $M_v
= -4 \cdots -5$, similar $M_v$s are expected for these SN Ibc progenitors.

\section{Relatively low-mass helium stars in binary systems}\label{sect:binary}

\begin{figure}
\centering
\includegraphics[width=1.0\columnwidth]{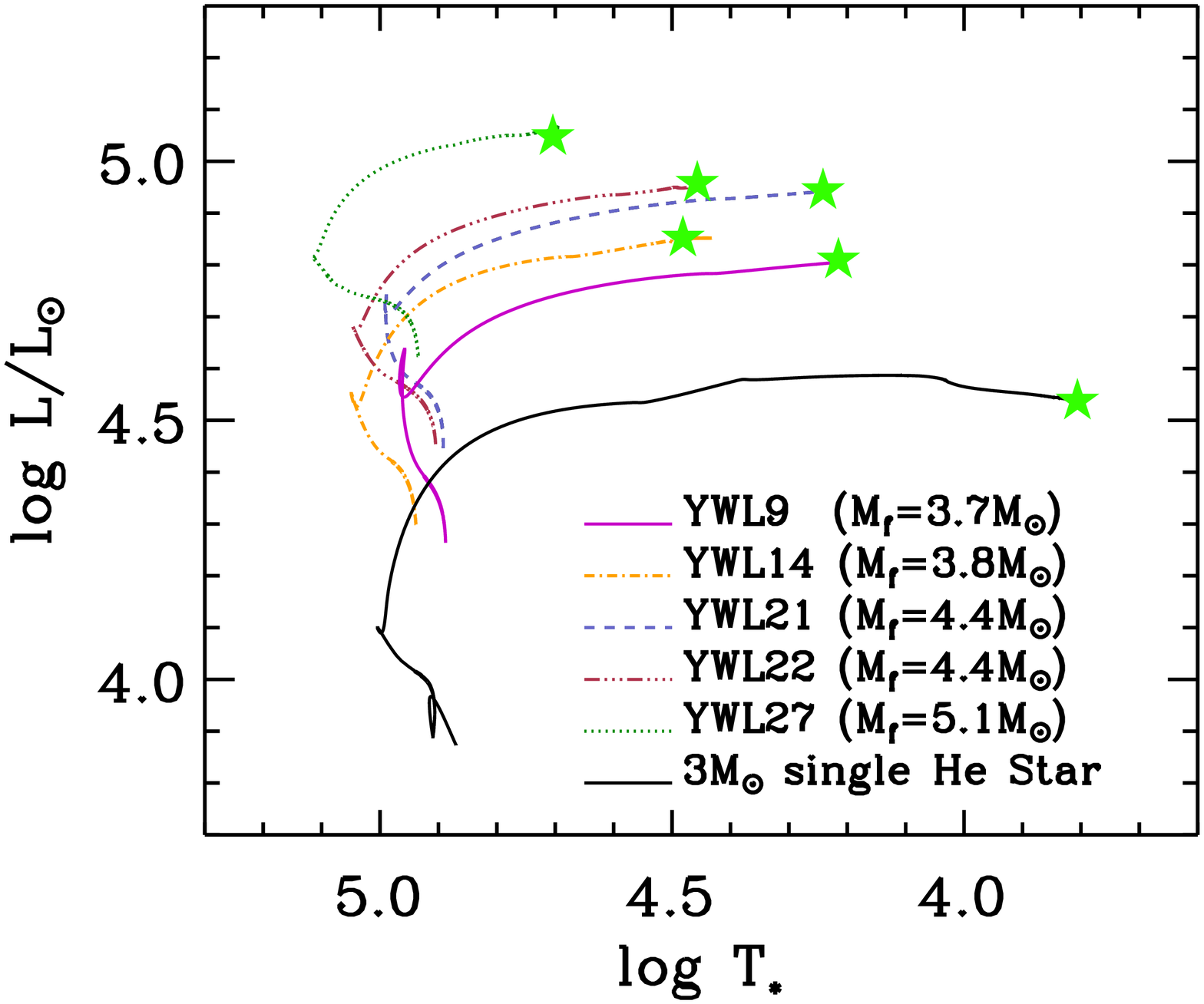}
\includegraphics[width=1.0\columnwidth]{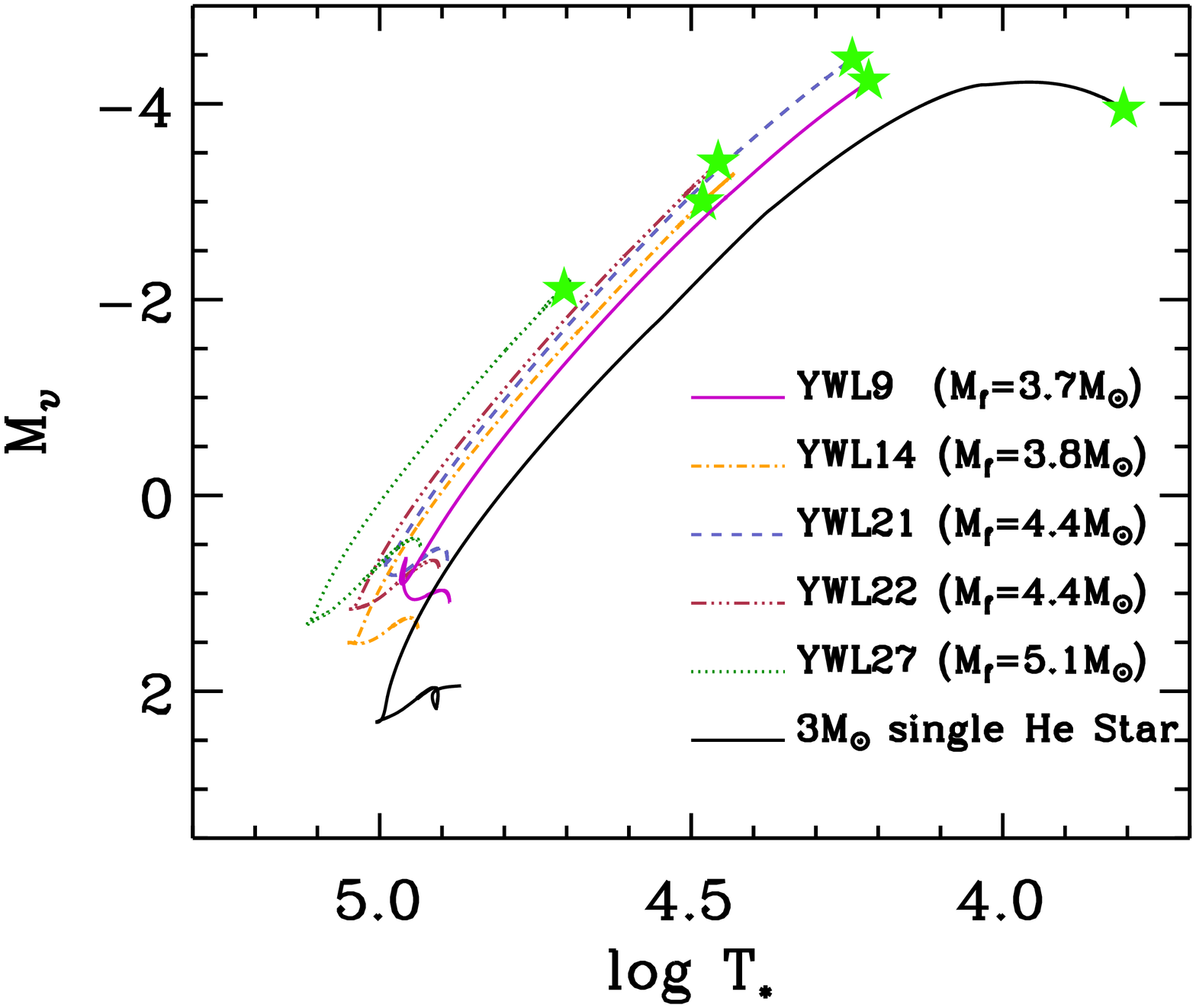}
\caption{\emph{Upper panel:} Evolutionary tracks of helium stars
produced in binary systems through
Case AB or B mass transfer phase, 
and of a single helium star of 3~\Msun{} (black solid line), at solar
metallicity. The star symbol marks the end point of each evolutionary track, 
which is the core carbon burning phase for 3~\Msun{} star
and the oxygen burning phase for the others.  
The helium-star models in binary systems are taken from
\citet{Yoon10}, and the evolutionary tracks are plotted only from the onset of
core helium burning.  The sequence number given for each track (except for the
3~\Msun{} single helium star) denotes the model sequence number in Table~1 of
\citet{Yoon10}.  The final mass ($M_f$) for each model sequence is given in the
parenthesis.  
\emph{Lower panel:} Corresponding absolute narrow-band visual magnitudes 
calculated with the assumption of black-body radiation. 
} 
\label{fig:hr_ywl} \end{figure}

Galactic WR stars have masses higher than about 8~\Msun{} and many of them are
thought to have originated from massive single stars through stellar-wind mass-loss. 
In binary systems, lower-mass helium stars can be produced via mass
transfer. As discussed by \citet{Yoon10}, such relatively low-mass helium stars
can retain a large amount of helium in the envelope and expand to a large
radius during the late evolutionary stages beyond core helium exhaustion.  The
final radii of SN Ibc progenitors are hence systematically larger for lower
masses.  

Fig.~\ref{fig:hr_ywl} shows the evolutionary tracks of several helium stars
produced in close binary systems taken from
\citet{Yoon10}.   Some of the helium stars in this figure 
(YWL9, YWL14 and YWL21) end their lives while they are filling
their Roche-lobes during Case ABB or BB mass transfer (i.e., mass transfer from the He star
during/after the carbon-oxygen core contraction phase; see
\citealt{Yoon10} for more details).  For comparison, the evolutionary track of a
single, pure helium star  of $M = 3$~\Msun{} is also presented in the figure.
Since these relatively low-mass helium stars may not have optically thick
winds, the black-body assumption is adopted as a rough approximation to
calculate the corresponding visual magnitudes, which are shown in
Fig.~\ref{fig:hr_ywl}.  
This figure shows that helium stars of $M = 3 M_\odot - 5M_\odot$ are visually very faint
($M_v > 1$) on the helium main-sequence. However,  rapid expansion of the 
helium envelope during and after the carbon-burning phase suddenly makes these
stars  luminous in the visual band shortly before their deaths.  In
particular, the surface temperature of 3~\Msun{} helium star can decrease to
$\log T_\star = 3.8$ and it  may be particularly luminous in the R band ($M_R \simeq
-5.5$). 

The important implication of this result is that some relatively 
low-mass SN Ibc progenitors ($\sim 3$~\Msun{}) are more luminous in either the
visual band ($M_v \simeq -4$) or other optical bands  than much more
massive SN Ibc progenitors with final masses higher than about 9~\Msun{},
which should be as dim as $M_v \simeq -1.5$ as discussed above (see also
Fig.~\ref{fig:mv}).

We note that not all SNe Ibc progenitors with final masses of
$M < 5$ \Msun{} may be optically bright.  Because they are likely to undergo
mass transfer in a close binary system during the final evolutionary stages owing
to the rapidly expanding envelopes,  it is possible that some of them may lose
most of their helium envelopes by the time of supernova explosion (if the
orbital separation is sufficiently short; ~\citealt{Pols02, Dewi02, Ivanova03,
Yoon10}).  These relatively low-mass, helium-deficient stars may constitute the
low-mass class of SN Ic progenitors~\citep{Pols02, Yoon10} and may explain the
fast light-curve of SN 1994I~\citep{Nomoto94}.  Such a SN Ic progenitor of the
low-mass class  has a surface temperature of $\log T_\star > 5.0$ and a very
faint optical luminosity ($M_v \gsim 1.0$).  On the other hand, helium stars of
about 2~\Msun{} -- 3~\Msun{} can also be produced by a merging of a white dwarf
and a helium star, in which case their surface temperatures may decrease below
$10^4$~K at the pre-supernova stage, which would appear to be optically bright
single stars. Therefore, fairly diverse properties are expected for SN Ibc
progenitor systems of this mass range.

\begin{table*}[ht]
\centering
\caption{Predicted visual magnitudes 
at the pre-supernova stage for different final masses of helium stars.}\label{tab1}
\begin{tabularx}{0.98\linewidth}{l l l l l}
\hline
Final Mass                         &   SN Type \tablefootmark{a}  & Helium envelope  & Single/Binary             &  $M_v$ \\
\hline
$M_f \gsim 9$~\Msun{}              &   Type Ic  & Stripped-off     & Single or Binary          &  $M_v = -1.5 \cdots -2.5$ \\
$ 6 \lsim M_f \lsim  8$~\Msun{}    &   Type Ib  & Retained         & Single or Binary          &  $ M_v = -4 \cdots -5$\tablefootmark{~b} \\
$ 2 \lsim M_f \lsim  5$~\Msun{}\tablefootmark{~c}    &   Type Ib  & Retained         & Binary or merger remnant  &  $ M_v = -2 \cdots -4$  ($M_R = -2.0 \cdots -5.5$) \\
$ 2 \lsim M_f \lsim  5 $~\Msun{}\tablefootmark{~c}   &   Type Ic  & Stripped-off     & Binary                    &  $  M_v \gsim 1 $ \\
\hline \\
\end{tabularx}
\tablefoot{
\tablefoottext{a}{ SN Ic is assumed if the helium envelope is stripped off from the progenitor ($M_\mathrm{He} \lsim 0.5$~\Msun) and
SN Ib otherwise. However, this may also depend on the history of chemical mixing during the supernova explosion as shown by \cite{Dessart12}.}
\tablefoottext{b}{This estimate is based on the assumption that the helium stars of this mass range still have optically thick WR winds. 
If not, $M_v$ is larger ($M_v \gsim -2$).} 
\tablefoottext{c}{The lower limit of 2~\Msun{}  is an arbitrary choice, but SNe Ibc with $M < 2$~\Msun{} are likely to be very faint and difficult to discover.}
}
\end{table*}

\section{Discussion}

Our study refutes the widespread belief that more massive
progenitors of SNe Ibc are more easily identifiable than less massive ones.
In contrast, our analysis implies that massive SN Ibc progenitors with
final masses  $M \gsim 9$~\Msun{} and bolometric luminosities  $\log
L/L_\odot \gsim 5.4$ may be optically faint compared to relatively low-mass SN
Ibc progenitors, depending on their mass-loss history.

The key factor that determines the brightness of a SN Ibc progenitor in optical
bands is the mass of the helium envelope. If most of the helium envelope  is
lost via either stellar winds or binary interactions, the SN Ibc progenitor
will be very hot, resulting in a low optical luminosity at the
pre-supernova stage.  This may be the case for the most massive WR stars as
discussed in Sect.~\ref{sect:wr}.  If a fairly thick helium envelope can be
retained in a SN Ibc progenitor, the expansion of the envelope that occurs
shortly before the SN explosion  makes the progenitor star fairly bright in
optical bands.  This case is mostly relevant for relatively low-mass helium
stars produced in close binary systems (Sect.~\ref{sect:binary}).
Therefore, we suggest that SN Ib progenitors would be optically brighter --
hence easier to detect -- than SN Ic progenitors, in general.  

We summarize in Table~\ref{tab1} our rough predictions for the visual magnitudes of SNe Ibc
progenitors for different possible cases.  A secure
identification of a SN Ic progenitor would be challenging.  
A detection limit of $M_v \simeq -1.5$ for SNe Ic of the high-mass class (i.e.,
$M_f \gsim 9$~\Msun) and a much better detection limit is needed for
SNe Ic of the low-mass class. Therefore, it is unsurprising that
\citet{Crockett07} could not find the progenitor star of SN Ic 2002ap
even with the best detection limit achieved so far (i.e., $M_B \gsim -4.2$ and
$M_R \gsim -5.1$).  Although SNe Ib progenitors are predicted to be
systematically more luminous than SNe Ic progenitors in optical bands, 
their visual luminosities would still be  limited to $M_v \simeq -5$ and a
wide range of optical magnitudes are expected.  

Given that many SNe Ibc are expected to occur in binary systems, another factor
that should be considered  in the search for SNe Ibc progenitors is the
companion star.  Even if a SN Ibc progenitor is optically faint, the companion
star may be detectable.  The binary star models indeed predict that  a
significant fraction of SNe Ibc progenitors  have an O-type companion.
However, not all of them are bright enough  to be identified with the detection
limits achieved in the previous observations. Our preliminary calculation
indicates that only a small fraction (less than about 10\%) of SNe Ibc
progenitors  have a luminous O star companion with $M_v < -4$.  

\section{Conclusion}
Our study has shown that the masses of SNe Ibc progenitor stars
are not linearly correlated with their optical brightness and
that the evolution of the surface properties of massive helium stars during their
final evolutionary stages should be carefully investigated.  The non-detection of a
SN Ibc progenitor even with a good detection limit does not
necessarily imply that its progenitor is a relatively low-mass helium star, and vice versa.  This should be
properly taken into account in future observational efforts to directly 
identify SNe Ibc progenitors.

\begin{acknowledgements}
We are grateful to  Georges Meynet, who referreed the paper, for his helpful comments. 
SCY would like to thank Norbert Langer for his support of this work. 
\end{acknowledgements}

\end{document}